\documentclass[a4paper]{jpconf}
\usepackage{graphicx}
\begin{document}
\title{Charm $CP$ violation and mixing at Belle}

\author{Byeong Rok Ko (on behalf of the Belle Collaboration)}

\address{Korea University, SEOUL 136-713, Republic of Korea}

\ead{brko@hep.korea.ac.kr}

\begin{abstract}
We present charm $CP$ violation and mixing measurements at Belle. They
are the first observation of $D^0-\bar{D}^0$ mixing in $e^+e^-$
collisions from $D^0\rightarrow K^+\pi^-$ decays, the most precise
mixing and indirect $CP$ violation parameters from $D^0\rightarrow
K^0_S\pi^+\pi^-$ decays, and the time-integrated $CP$ asymmetries in
$D^0\rightarrow\pi^0\pi^0$ and $D^0\rightarrow K^0_S\pi^0$ decays. Our
mixing measurement in $D^0\rightarrow K^+\pi^-$ decays excludes the
no-mixing hypothesis at the 5.1 standard deviation level. The mixing
parameters $x=(0.56\pm0.19^{+0.03+0.06}_{-0.09-0.09})\%$,
$y=(0.30\pm0.15^{+0.04+0.03}_{-0.05-0.06})\%$ and indirect $CP$
violation parameters
$|q/p|=(0.90^{+0.16+0.05+0.06}_{-0.15-0.04-0.05})$,
arg$(q/p)=(-6\pm11\pm3^{+3}_{-4})^{\circ}$ measured from
$D^0\rightarrow K^0_S\pi^+\pi^-$ decays, and the time-integrated $CP$
asymmetries
$A^{D^0\rightarrow\pi^0\pi^0}_{CP}=(-0.03\pm0.64\pm0.10)\%$ and
$A^{D^0\rightarrow K^0_S\pi^0}_{CP}=(-0.21\pm0.16\pm0.07)\%$ are the
most precise measurements to date. Our measurements here are
consistent with predictions of the standard model.
\end{abstract}

The magnitudes of $CP$ violation ($CPV$) and mixing rate in the charm
system are very small in the standard model
(SM)~\cite{SM1,SM2}. Therefore, $D^0$-$\bar{D}^0$ mixing and $CPV$
measurements provide a unique probe to search for physics beyond the
SM~\cite{BSM1,BSM2}. 
$D^0-\bar{D}^0$ mixing occurs since the mass eigenstates $D_1$ and
$D_2$ are different from the flavor eigenstates $D^0$ and
$\bar{D}^0$. The mass eigenstates can be written in terms of the
flavor eigenstates, namely, $|D_{1,2}\rangle=p|D^0\rangle\pm
q|\bar{D}^0\rangle$. The phenomenology of meson mixing is described by
two parameters, $x=\Delta m/\Gamma$ and $y=\Delta\Gamma/2\Gamma$,
where $\Delta m$ and $\Delta\Gamma$ are the mass and width differences
between the two mass eigenstates, and $\Gamma$ is the average decay
width of the mass eigenstates. Indirect $CPV$ parameters are $|q/p|$
and arg$(q/p)$, where the former and the latter are responsible for
$CPV$ in mixing and that in interference between the decays with and
without mixing, respectively. 

In this proceedings, we present charm $CPV$ and mixing measurements
from the data recorded with the Belle detector~\cite{Belle_detector}
at the $e^+e^-$ asymmetric-energy collider KEKB~\cite{KEKB}.

\section{Mixing in $D^0\rightarrow K^+\pi^-$ decays}
\begin{figure}[h]
\begin{minipage}{18pc}
\includegraphics[width=18pc]{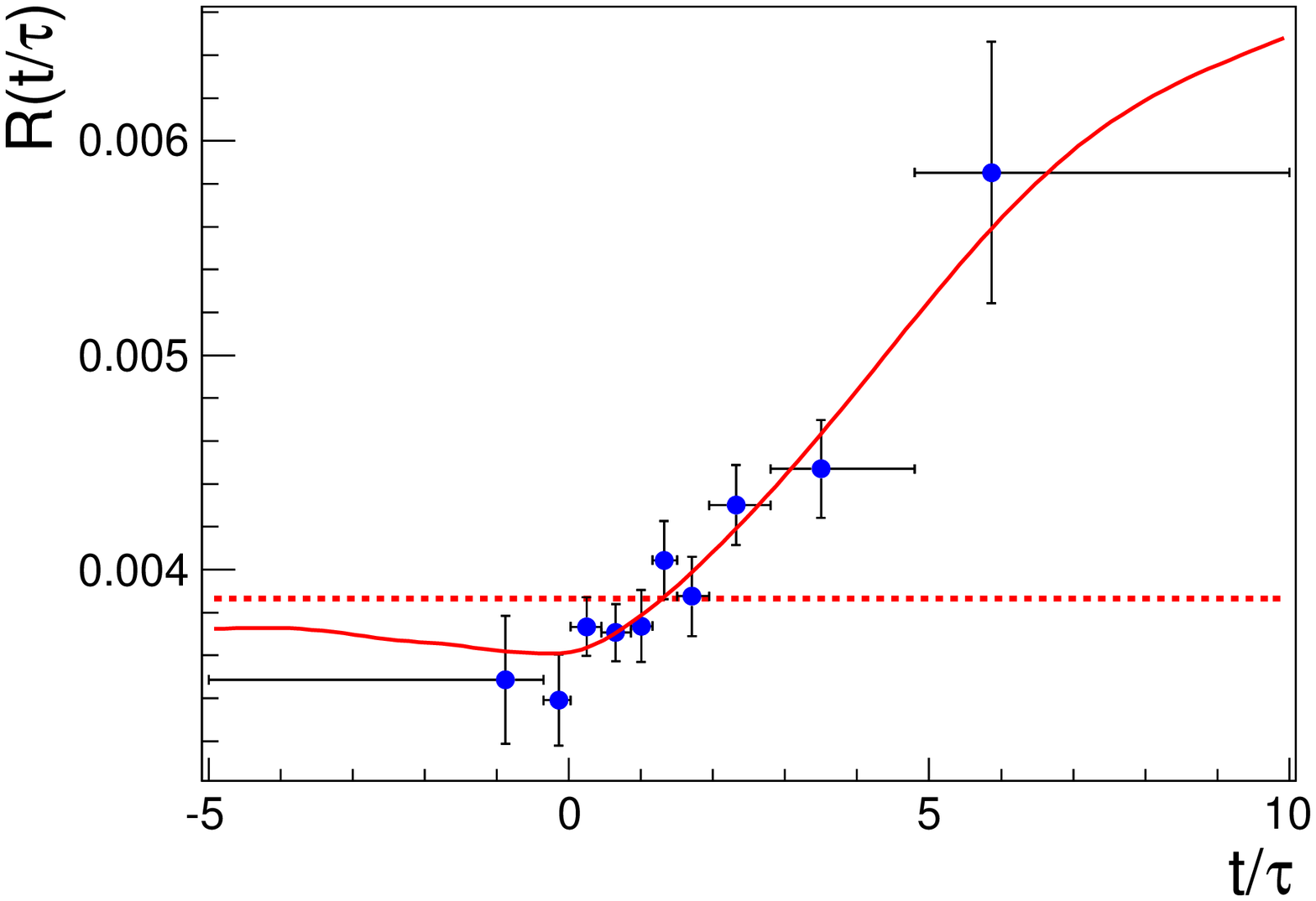}
\caption{\label{mixing_Rt}The time-dependent ratios of WS to RS decay rates. Points
    with error bars reflect the data and their total
    uncertainties. The lines show the fit with (solid) and without
    (dashed) the mixing hypothesis.}
\end{minipage}\hspace{2pc}%
\begin{minipage}{18pc}
\includegraphics[width=18pc]{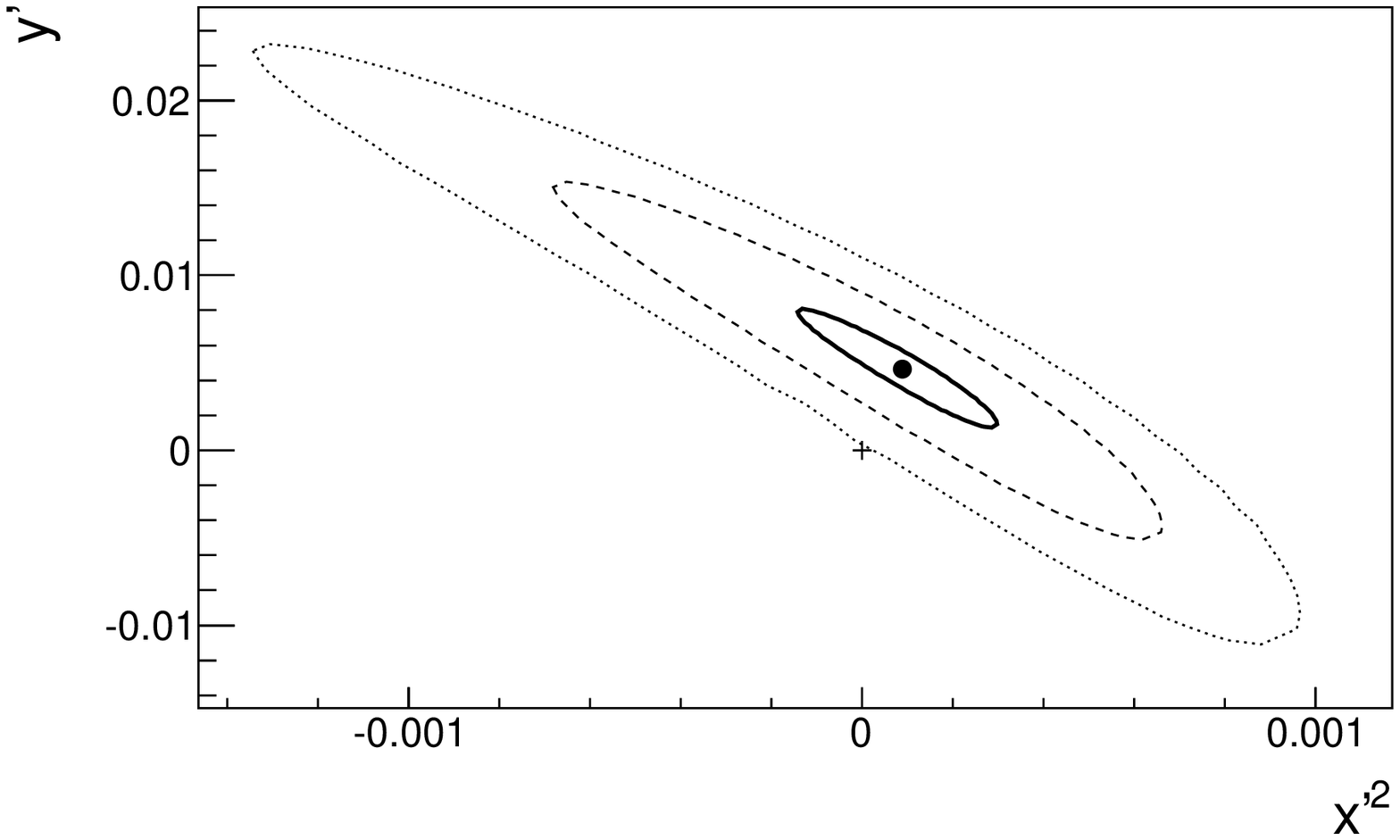}
\caption{\label{mixing_x2y}Best-fit point and contours in the $({x'}^2,y')$ plane. The
    solid, dashed, and dotted lines, respectively, correspond to 1, 3,
    and 5 standard Gaussian deviations from the best fit. The cross is
    the no-mixing point.}
\end{minipage}
\end{figure}

We present the first observation of $D^0-\bar{D}^0$ mixing from an
$e^+e^-$ collision experiment by measuring the time-dependent ratio of
the $D^0\to K^+\pi^-$ ({\bf W}rong {\bf S}ign) to $D^0\to K^-\pi^+$
({\bf R}ight {\bf S}ign) decay rates~\cite{Belle_WS}. Assuming $CP$
conservation and that the mixing parameters are small ($|x|\ll 1$ and
$|y|\ll 1$), the time-dependent RS and WS decay rates are
$\Gamma_{\rm RS}(\tilde{t}/\tau)\approx|\mathcal{A}_{\rm CF}|^2
e^{-\frac{\tilde{t}}{\tau}}$ and $\Gamma_{\rm
  WS}(\tilde{t}/\tau)\approx|\mathcal{A}_{\rm CF}|^2
e^{-\frac{\tilde{t}}{\tau}}\Biggl(R_D +
\sqrt{R_D}y'\frac{\tilde{t}}{\tau} +
\frac{{x'}^2+y'^2}{4}\biggl(\frac{\tilde{t}}{\tau}\biggr)^2\Biggr)$,
respectively, to second order in the mixing parameters, where
$\tilde{t}$ is the true proper decay time, $\mathcal{A}_{\rm CF}$ is
the {\bf C}abibbo-{\bf F}avored decay amplitude, $\tau$ is the $D^0$
lifetime, $R_D$ is the ratio of DCS ({\bf D}oubly {\bf C}abibbo-{\bf
  S}uppressed) to CF decay rates, $x'=x\cos\delta+y\sin\delta$, and
$y'=y\cos\delta-x\sin\delta$, where $\delta$ is the strong phase
difference between the DCS and CF decay amplitudes. The time-dependent
ratio of WS to RS decay rates then
\begin{equation}
  R(t/\tau)=\frac{\int^{+\infty}_{-\infty}\Gamma_{\rm WS}(\tilde{t}/\tau)\mathcal{R}(t/\tau-\tilde{t}/\tau)d(\tilde{t}/\tau)}{\int^{+\infty}_{-\infty}\Gamma_{\rm RS}(\tilde{t}/\tau)\mathcal{R}(t/\tau-\tilde{t}/\tau)d(\tilde{t}/\tau)},
  \label{EQ:RWS_EXP}
\end{equation}
where $t$ is the reconstructed proper decay time and
$\mathcal{R}(t/\tau-\tilde{t}/\tau)$ is the resolution function of the
real decay time, $\tilde{t}$. Figure~\ref{mixing_Rt} shows the
time-dependent ratios of WS to RS decays together with the two
hypothesis tests, with (line) and without mixing (dots). The $\chi^2$
difference between the ``no-mixing'' and ``mixing'' hypotheses,
$\Delta\chi^2=\chi^2_{\rm no-mixing}-\chi^2_{\rm mixing}$, is 29.3 for
two degrees of freedom, corresponding to a probability of
4.3$\times10^{-7}$; this implies the no-mixing hypothesis is excluded
at the 5.1 standard deviation level. Thus, we observe $D^0-\bar{D}^0$
mixing for the first time in an $e^+e^-$ collision experiment. We also
show this in Figure~\ref{mixing_x2y} with the 1$\sigma$, 3$\sigma$,
and 5$\sigma$ contours around the best fit point in the $({x'}^2,y')$
plane.

\section{Mixing and indirect $CPV$ in $D^0\rightarrow K^0_S\pi^+\pi^-$ decays}
\begin{figure}[h]
\begin{minipage}{18pc}
\includegraphics[width=18pc]{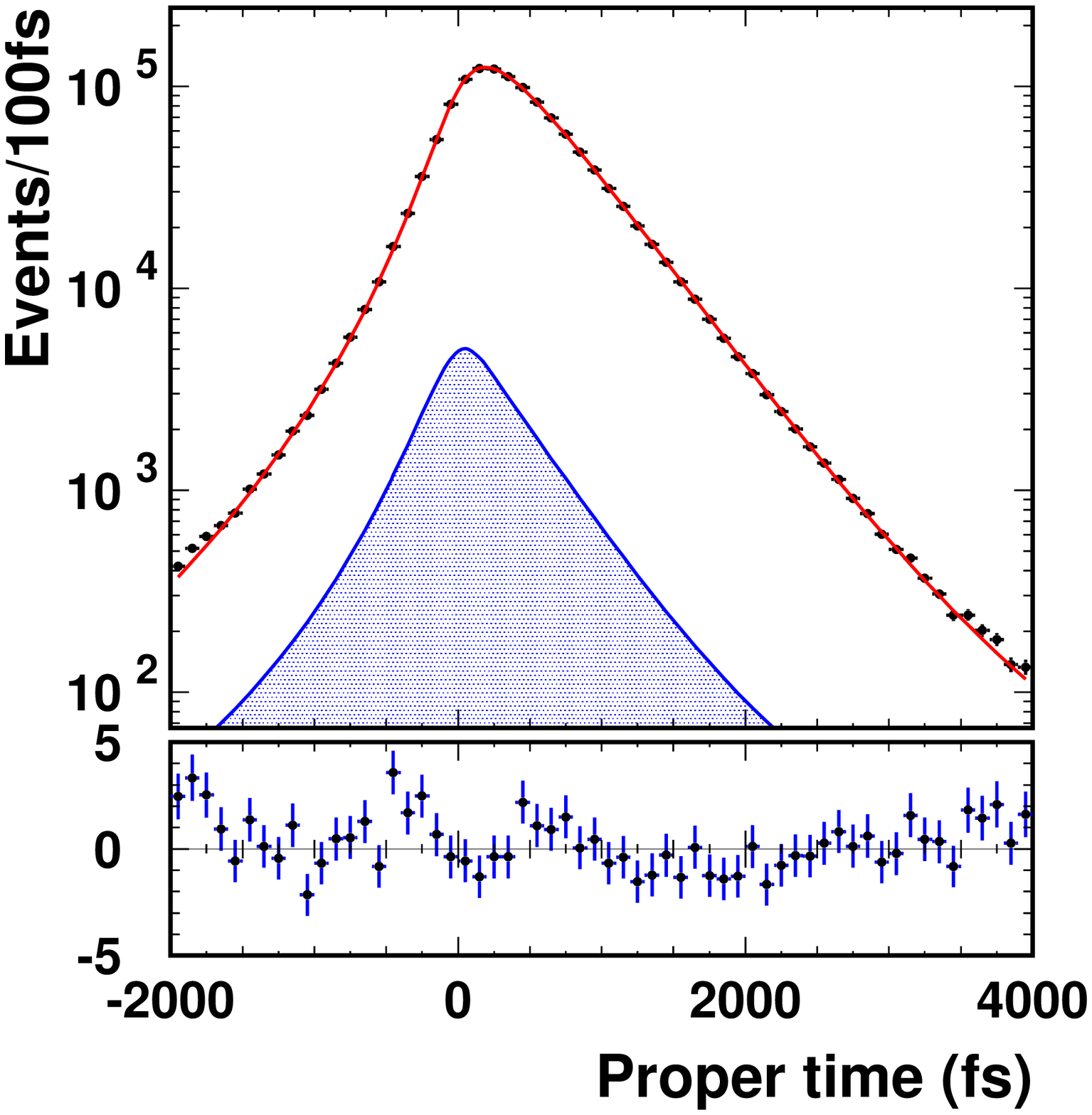}
\caption{\label{mixing_dtime}The proper-time distribution for events in the
signal region (points) and fit projection for the $CP$ conserved fit
(curve). The shaded region shows the combinatorial components. The
residuals are shown below the plot.}
\end{minipage}\hspace{2pc}%
\begin{minipage}{18pc}
\includegraphics[width=18pc]{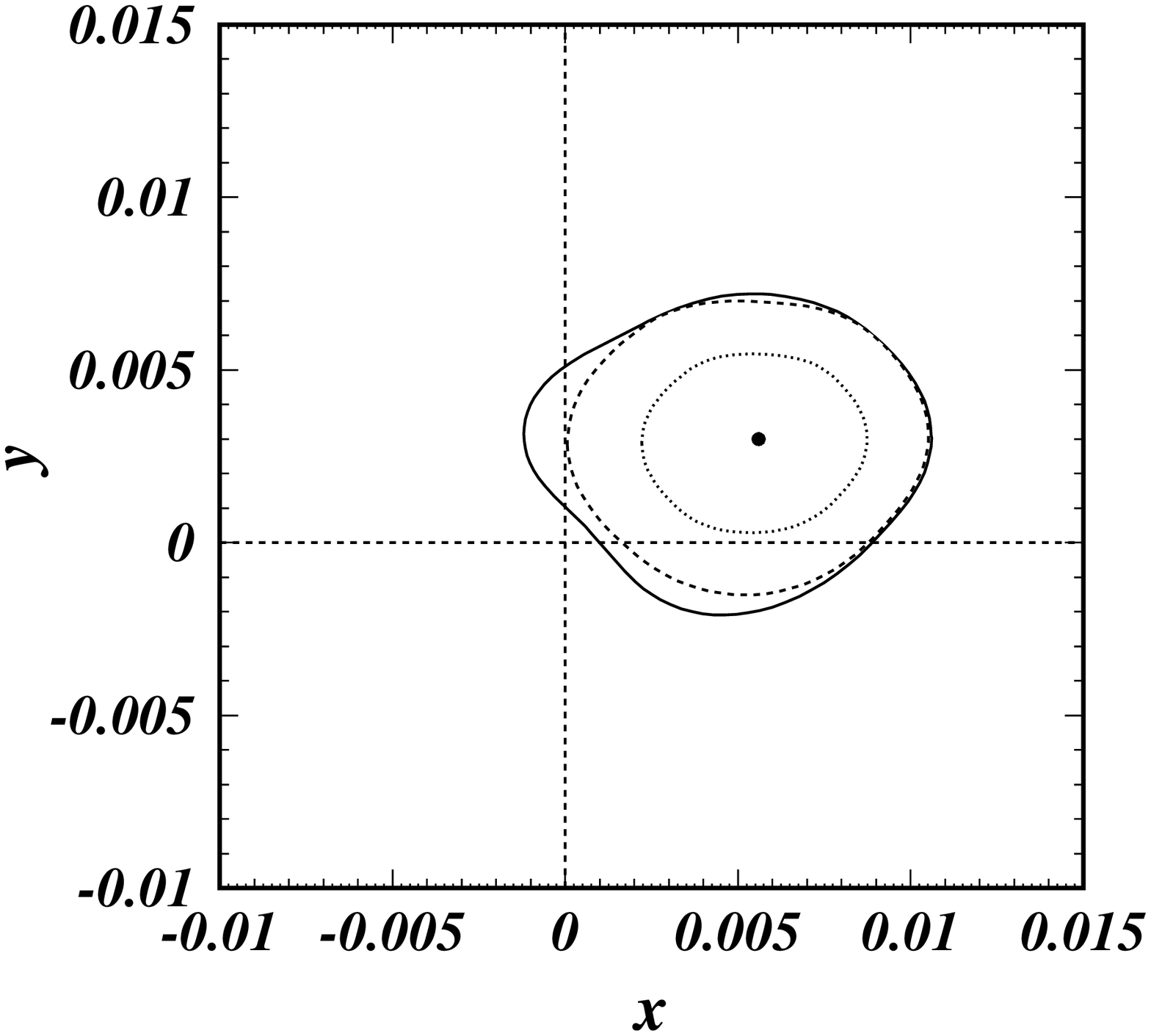}
\caption{\label{mixing_xy}Central value (point) and C.L. contours for $(x,
y)$ : dotted (dashed) corresponds to 68.3\% (95\%) C.L. contour for
$CP$-conserved Dalitz fit, and solid corresponds to 95\% C.L. contour
for $CPV$-allowed fit with statistical, experimental and model
uncertainties included.}
\end{minipage}
\end{figure}

We present the most precise mixing and indirect $CP$ violation
parameters from $D^0\rightarrow K^0_S\pi^+\pi^-$
decays~\cite{Belle_kspipi} using the time-dependent Dalitz fit
analysis~\cite{CLEO}. The time-dependent decay matrix elements of
$D^0\rightarrow K^0_S\pi^+\pi^-$ and $\bar{D}^0\rightarrow
K^0_S\pi^+\pi^-$ are
\begin{eqnarray}
  \nonumber
  \mathcal{M}(m^2_+,m^2_-,t)&=&g_+(t)\mathcal{A}(m^2_+,m^2_-)+\frac{q}{p}g_-(t)\bar{\mathcal{A}}(m^2_+,m^2_-)~{\rm and}\\
  \bar{\mathcal{M}}(m^2_+,m^2_-,t)&=&g_+(t)\bar{\mathcal{A}}(m^2_+,m^2_-)+\frac{p}{q}g_-(t)\mathcal{A}(m^2_+,m^2_-),
  \label{EQ:ME}
\end{eqnarray}
where $m^2_{\pm}=m^2_{K^0_S\pi^{\pm}}$, $\mathcal{A}(m^2_+,m^2_-)=\sum
a_j e^{i\delta_i}A_j(m^2_+,m^2_-)$,
$\bar{\mathcal{A}}(m^2_+,m^2_-)=\sum \bar{a}_j
e^{i\bar{\delta}_i}A_j(m^2_+,m^2_-)$, $g_{\pm}(t)=(e^{-i\lambda_1
  t}\pm e^{-i\lambda_2 t})/2$, and $\lambda_i=m_i-i\Gamma_i/2$, where
$a_j$, $\delta_j$, and $A_j(m^2_+,m^2_-)$ are amplitude, phase, and
resonance matrix element, respectively. The decay matrix element
squared $\mathcal{M}^2$ then contains the mixing parameters $x$ and
$y$ as well as $q/p$ of which magnitude and argument are indirect
$CPV$ parameters. Two separate time-integrated Dalitz fits to $D^0$
and $\bar{D}^0$ samples show no direct $CPV$ resulting in
$a_j\approx\bar{a}_j$ and $\delta\approx\bar{\delta}$. Hence, we
search for indirect $CPV$ assuming no direct $CPV$, namely,
$\mathcal{A}(m^2_+,m^2_-)=\bar{\mathcal{A}}(m^2_+,m^2_-)$. The best
fit model for $\mathcal{A}(m^2_+,m^2_-)$ is found to be a sum of
twelve Breit-Wigner for $P$- and $D$-wave resonances,
K-matrix~\cite{KMATRIX} and LASS~\cite{LASS} models for $\pi\pi$ and
$K\pi$ S-wave states, respectively, without non-resonant
decay. Figure~\ref{mixing_dtime} shows the proper-time distribution
superimposing the fit under both direct and indirect $CP$ conservation
and the fit results are $x=(0.56\pm0.19^{+0.03+0.06}_{-0.09-0.09})\%$
and $y=(0.30\pm0.15^{+0.04+0.03}_{-0.05-0.06})\%$. With allowing
indirect $CPV$, the fit returns
$x=(0.56\pm0.19^{+0.04+0.06}_{-0.08-0.08})\%$,
$y=(0.30\pm0.15^{+0.04+0.03}_{-0.05-0.07})\%$,
$|q/p|=(0.90^{+0.16+0.05+0.06}_{-0.15-0.04-0.05})$, and
arg$(q/p)=(-6\pm11\pm3^{+3}_{-4})^{\circ}$. Our measurements of
indirect $CPV$ parameters are the most accurate to date, but
consistent with no $CPV$. Figure~\ref{mixing_xy} show the
two-dimensional $(x,y)$ confidence-level (C.L.) contours for the
$CP$-conserved and $CPV$-allowed fits, where the no-mixing point
$(x=0,y=0)$ is excluded with 2.5 standard deviations.

\section{Time-integrated $CPV$ in $D^0\rightarrow\pi^0\pi^0$ and $D^0\rightarrow K^0_S\pi^0$ decays}
\begin{figure}[h]
\begin{minipage}{36pc}
\includegraphics[width=36pc]{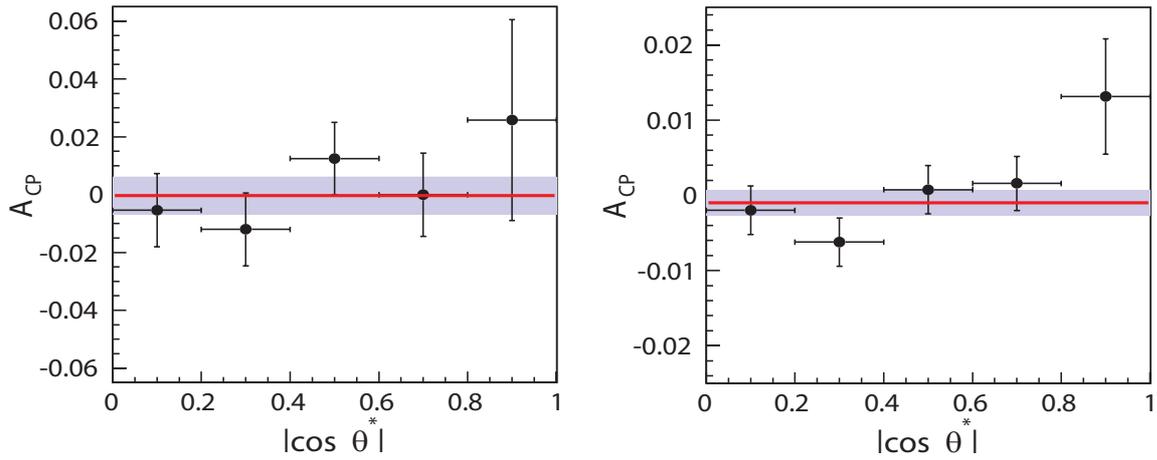}
\caption{\label{acp_pipi} The $A_{CP}$ values in the decays
  $D^0\rightarrow\pi^0\pi^0$ (left) and $D^0\rightarrow K^0_S\pi^0$
  (right) as a function of $|\cos\theta^*|$. The solid lines and
  shaded regions represent the central value and 1$\sigma$ interval of
  the $A_{CP}$.}
\end{minipage}
\end{figure}

We present the time-integrated $CPV$ asymmetry in $D^0\to f$,
$A^{D^0\to f}_{CP}$, where $f$ is the final state $\pi^0\pi^0$ or
$K^0_S\pi^0$~\cite{Belle_2pi0}. It is measured through the
reconstruction asymmetry
\begin{equation}
A_{\rm rec}=\frac{N_{\rm rec}^{D^{*+}\to D^0\pi^+_{s}}-N_{\rm rec}^{D^{*-}\to\bar{D}^0\pi^-_{s}}}
{N_{\rm rec}^{D^{*+}\to D^0\pi^+_{s}}+N_{\rm
    rec}^{D^{*-}\to\bar{D}^0\pi^-_{s}}}\approx A^{D^0\to f}_{CP} +
A^{D^{*+}}_{FB} + A^{\pi^+_s}_{\epsilon} (+ A^{\bar{K}^0}_{\epsilon}),
\label{EQ:ASYM_REC}
\end{equation} 
where $N_{\rm rec}$ is the number of reconstructed signal events,
$A_{FB}$ is the forward-backward asymmetry, $A^{\pi^+_s}_{\epsilon}$
is the detection asymmetry between positively and negatively charged
soft pions, and $A^{\bar{K}^0}_{\epsilon}$ is the asymmetry due to
different nuclear interactions between $\bar{K}^0$ and
$K^0$~\cite{KOMAT}, thus included to the decay $D^0\to K^0_S\pi^0$.
Once we remove $A^{\pi^+_s}_{\epsilon}$ with the correction in
Ref.~\cite{Belle_ksp0} and $A^{\bar{K}^0}_{\epsilon}$ with the value
in Ref.~\cite{Belle_kspi}, then $A_{\rm rec}$ becomes to have
$A^{D^0\to f}_{CP}$ and $A^{D^{*+}}_{FB}$ only, where the latter is an
odd function of the cosine of the polar angle of the $D^{*+}$ momentum
in the center-of-mass system ($\cos\theta^{*}_{D^{*+}}$). Using the
antisymmetry of $A^{D^{*+}}_{FB}$ in $\cos\theta^{*}_{D^{*+}}$, we
obtain $A_{CP}$ as a function of $|\cos\theta^{*}_{D^{*+}}|$ as shown
in Figure~\ref{acp_pipi}. The central $A_{CP}$ values obtained from a
least-square minimization are
$A^{D^0\rightarrow\pi^0\pi^0}_{CP}=(-0.03\pm0.64\pm0.10)\%$ and
$A^{D^0\rightarrow K^0_S\pi^0}_{CP}=(-0.21\pm0.16\pm0.07)\%$ which are
the most precise measurements to date revealing no $CPV$, thus
consistent with the SM.

\ack{B. R. Ko acknowledges support by NRF Grant No. NRF-2014R1A2A2A01005286.}

\end{document}